\documentclass[preprint]{elsarticle}
\usepackage{amsmath}
\usepackage{amsfonts}
\usepackage{gensymb}
\usepackage{lineno,hyperref}
\usepackage[dvipsnames]{xcolor}
\usepackage{subcaption}
\usepackage{xcolor}
\usepackage[utf8x]{inputenc}
\usepackage{chngcntr}
\usepackage{enumitem}
\counterwithout{figure}{subsection}
\usepackage[final]{changes}
\usepackage{mathtools}
\usepackage[linesnumbered, ruled, vlined]{algorithm2e}
\usepackage{siunitx}
\ExplSyntaxOn
\cs_set_eq:NN \__siunitx_unit_output_number_sep:
  \__siunitx_unit_output_number_sep_aux:
\ExplSyntaxOff
\DeclareSIUnit{\Ci}{\text{Ci}}
\DeclareSIUnit{\MeVee}{\text{MeVee}}
\journal{Nuclear Inst. and Methods in Physics Research, A}

\makeatletter
\def\ps@pprintTitle{
 \def\@oddfoot{}%
 \let\@evenfoot\@oddfoot}
\makeatother

\begin{document}


\begin{frontmatter}

\title{Enabling pulse shape discrimination with commercial ASICs}


\author[mymainaddress]{John Leland}

\author[mymainaddress]{Ming Fang}
\author[mymainaddress,paniaddress]{Satwik Pani}
\author[mymainaddress1]{Yuri Venturini}
\author[mymainaddress1]{Marco Locatelli}
\author[mymainaddress]{Angela Di Fulvio\corref{mycorrespondingauthor}}
\cortext[mycorrespondingauthor]{Corresponding author. Tel.: +1 217 300 3769.}
\ead{difulvio@illinois.edu}

\address[mymainaddress]{Department of Nuclear, Plasma, and Radiological
                        Engineering, University of Illinois, Urbana-Champaign,
                        104 South Wright Street, Urbana, IL 61801, United
                        States}
\address[paniaddress]{Department of Radiation Oncology, Washington University School of Medicine, 660 S. Euclid Ave, St. Louis, MO, 63110 USA}
\address[mymainaddress1]{CAEN Technologies Inc., 1 Edgewater Street - Suite 101
Staten Island, NY 10305, United States}


\begin{abstract}
        
        Fast electronic readout for high-channel density scintillator-based systems is needed for radiation tracking and imaging in a wide range of applications, including nuclear physics, nuclear security and nonproliferation. 
        Programmable electronics, like FPGAs and ASICs, provide a fast way of conditioning and processing the signal in real time. In this paper, we present a pulse shape discrimination (PSD) method based on the shaping circuit of a commercially available ASIC, the Citiroc1A by CAEN Technologies. We used two different shaping times per detector channel to calculate a shaping parameter that enables PSD. Using our new method, neutron and gamma-ray pulses detected by a d$_{12}$-stilbene scintillator can be effectively discriminated at light output values greater than \SI{0.15}{\MeVee}. While not achieving the PSD performance of traditional offline charge integration, our method does not require the transfer of data to a separate system for further processing and enables the direct deployment of high-channel density multi-particle detection systems. Moreover, the availability of a wider range of shaping times than those on the Citiroc1A can potentially further improve the PSD performance.
        
        
\end{abstract}

\begin{keyword}
{silicon photomultiplier, pulse shape discrimination, application specific integrated circuit}
\end{keyword}

\end{frontmatter}

\section{Background and Motivation}
The availability of solid-state light readout devices, like silicon photomultipliers (SiPMs), makes it possible to scale large arrays of scintillation detectors~\cite{goldsmithGerling2016,5168050} to hand-held versions~\cite{osti_1172910,single_vol_scatter}. Compared to avalanche photo-detectors and photo-diodes, SiPMs exhibit better performances, e.g., higher gain, lower noise, and a faster response~\cite{PIEMONTE20192}. When SiPMs are coupled with pulse-shape discrimination (PSD) capable scintillators, they allow for the simultaneous detection of different types of radiation, e.g., gamma rays and neutrons.  Furthermore, the advantage of using SiPMs is the ability to handle a high number of channels\added{, in the order of hundreds,} while maintaining a compact form factor, ideal for hand-held stand-off gamma ray and neutron imaging. Programmable electronics, like field programmable gate arrays (FPGA) and application specific integrated circuits (ASICs), are particularly suitable for the readout of SiPMs coupled to arrays of scintillation detectors because they can perform similar complex functions as traditional analog electronics~\cite{costrell1970development} and fast digitizers~\cite{KORNILOV2003467} while being compact and featuring a high number of channels. However, implementing specific functions, such as PSD, requires the design of custom ASIC-based devices, which are cost prohibitive for small-scale projects. 

In this work, we address this issue by developing a method to perform PSD using the commercially available Citiroc1A ASIC controlled by the A55CIT4-DT5550W readout system~\cite{a55citx2020,dt5550w2020} by CAEN Technologies (Viareggio, Italy). This PSD method does not require sampling and storing hundreds of samples for each detected pulse, which corresponds to the analog-to-digital conversion process implemented on fast digitizers. Additionally, our method can be performed on-board without the need of transferring the data to another device for analysis.

The ASIC-based method implements an architecture that extracts a pulse-shape-dependent parameter, which enables the discrimination of gamma-ray and neutron pulses. In the 70s, similar analog PSD methods were explored and implemented. These methods typically perform PSD using a shaping stage followed by a time-to-amplitude circuit and comparator that discriminates between neutron and gamma-ray pulses \cite{KINBARA1969173} \cite{HEISTEK1970213}.
While being inspired by the analog implementation, our method expands on the prior work by translating the analog architectures to digital processing and enables PSD for high-channel density detector arrays using advanced on-board signal processing. Therefore, the method we propose grants the capability to perform PSD on a compact detection system with a high number of channels for a broad range of applications, including nonproliferation, nuclear physics, and radiation protection.


We developed and implemented an on-board PSD method for the commercially available Citiroc1A ASIC~\cite{citiroc1a}. In Section~\ref{sec:methods}, we introduce the ASIC-based method, which we will call Digital Shaper (DS) PSD, to discriminate between neutron and gamma-ray pulses and calculate the circuit response function for optimizing the PSD parameter using template pulses. In Section~\ref{sec:results}, we apply the optimized PSD method on measured gamma-ray and neutron pulses. A comparison between the traditional charge-integration (CI) PSD and DS PSD is given. The discussion and conclusions are presented in Section~\ref{sec:conclusion}
\section{Methods}\label{sec:methods}
In this section, we describe the implementation of PSD using electronics available in the Citiroc1A ASIC. The DS PSD method will be demonstrated on synthetic pulses and experimentally measured data, and its performance will be compared with the traditional PSD method.
\subsection{Imaging System Design}
\begin{figure}[!htbp]
    \centering
    \includegraphics[width=\linewidth]{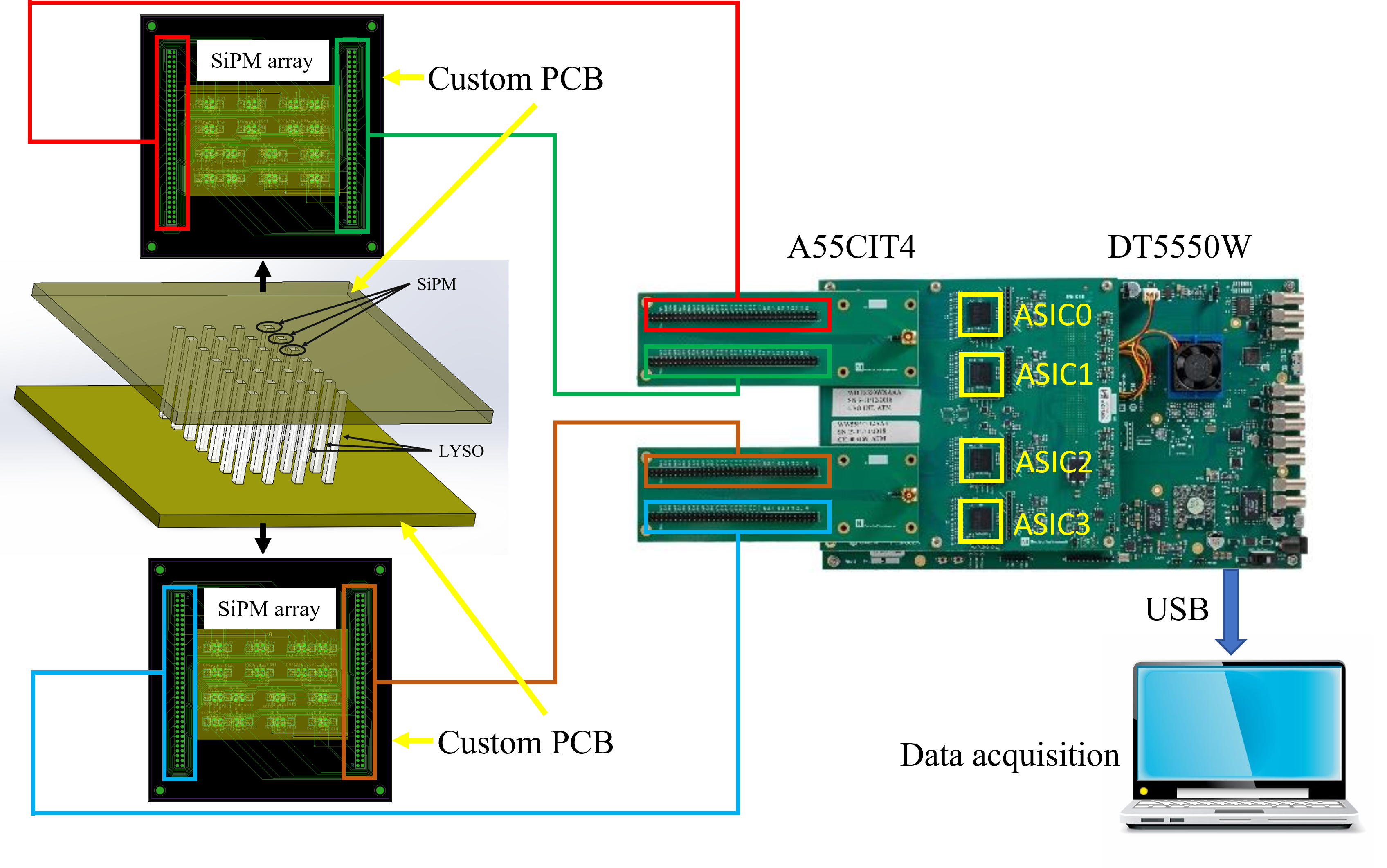}
    \caption{Schematic of a \added{multi-particle spectroscopic imager} \replaced{based on}{using} an ASIC-based electronic readout.}
    \label{fig:design}
\end{figure}
We built a compact imaging system encompassing a $4\times7$ array of LYSO crystals (\SI{3}{\mm}$\times$\SI{3}{\mm}$\times$\SI{5}{\cm}) coupled to 56 Onsemi MicroFJ 30020 SiPMs~\cite{sensljseries2015} that are read out by four Citiroc1A ASICs. In the future the LYSO will be replaced by PSD-capable scintillators such as stilbene. This detector can be used as an imager for nuclear security and non-proliferation applications or as a particle tracker for nuclear physics experiments. Figure~\ref{fig:design} shows the schematic of the imager. The system includes two custom printed circuit boards (PCBs), each powering and controlling the readout of a $4\times7$ array of SiPMs. The SiPM arrays are used for light readout of the top and bottom sides of a $4\times7$ scintillator matrix. The dual-end readout enables the retrieval of \replaced{depth of interaction information~\cite{morrocchi2017depth,ruch2016position}.}{depth information of the interactions.} The SiPM pulses are fed to an A55CIT4-DT5550W readout system by CAEN Technologies~\cite{dt5550w2020}. The system is controlled via computer using the DT5550W Readout Software (Sci Digitizer Family, Version 2022.1.1.0, distributed by CAEN Technologies). This software allows us to control the DT5550W and program some of the board settings, including the gain, the threshold, and the shaping constants. The A55CIT4 board hosts four Citiroc1A ASICs, each with a 32 readout channel capacity~\cite{a55citx2020}. Thus, there are 128 input channels available in total. This board is particularly suitable for SiPM readout because it provides a bias from \SI{20}{\V} to \SI{85}{\V} with a current up to \SI{10}{\mA}~\cite{A7585D}. In the current configuration, the board is set to provide a \SI{30}{\V} bias to the SiPMs. The A55CIT4 board features a fine time resolution better than \SI{100}{\ps} as stated by the manufacturer~\cite{a55citx2020} and a 14-bit ADC with a sensitivity of 160 fC.
We tested the effect of the Citiroc1A's dead time on our data throughput using the DT5810B pulse generator. We found no loss of data when up to a \SI{10}{\kHz} input frequency. This acquisition rate is compatible with the detection of an approximately \SI{6}{m\Ci} source at a distance of \SI{1}{\m} with our imager without. Therefore, the proposed configuration is suitable for imaging applications, even when measuring relatively high-intensity sources.
\subsection{PSD Method}\label{sec:PSD_Method}
In this section, we briefly summarize the principles of two PSD methods, namely the traditional charge-integration PSD and the proposed DS PSD.\added{ The datasets used for the evaluation of the PSD methodologies are:}
\begin{enumerate}
    \item[A.] Synthetic dataset acquired using a pulse generator connected to the ASIC
    \item[B.] Measured dataset produced using the SiPM evaluation board, digitizer, and stilbene-d$_{12}$
    \item[C.] Measured dataset produced using SiPM evaluation board, ASIC, and stilbene-d$_{12}$
\end{enumerate}
\subsubsection{Charge-integration PSD}
\begin{figure}
    \centering
    \begin{subfigure}{.5\textwidth}
        \includegraphics[width=\textwidth]{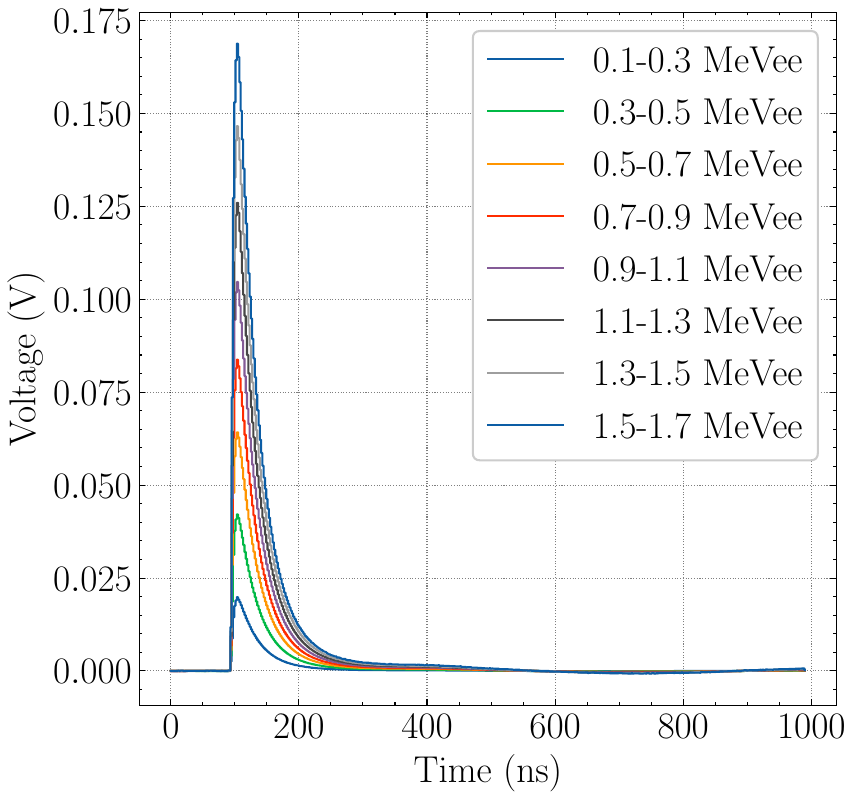} 
        \caption{\added{Average gamma-ray pulses.}}
        \label{fig:gamma_template_LO_binned}
    \end{subfigure}%
    \begin{subfigure}{.5\linewidth}
        \includegraphics[width=\textwidth]{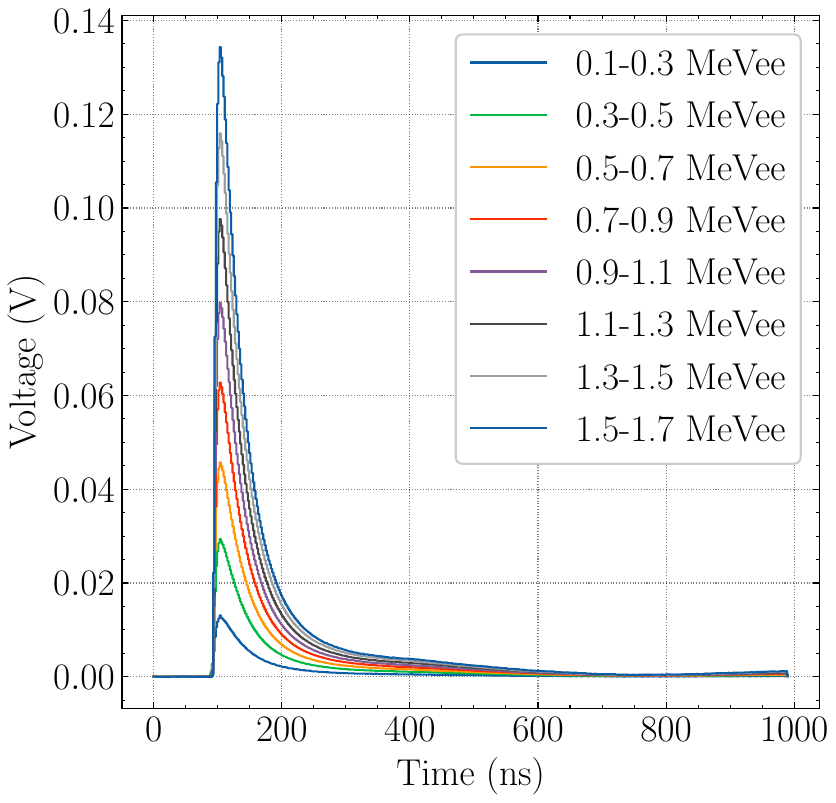} 
        \caption{\added{Average neutron pulses.}}
        \label{fig:neutron_template_LO_binned}
    \end{subfigure}\\[1ex]
        \begin{subfigure}{.5\textwidth}
        \includegraphics[width=\textwidth]{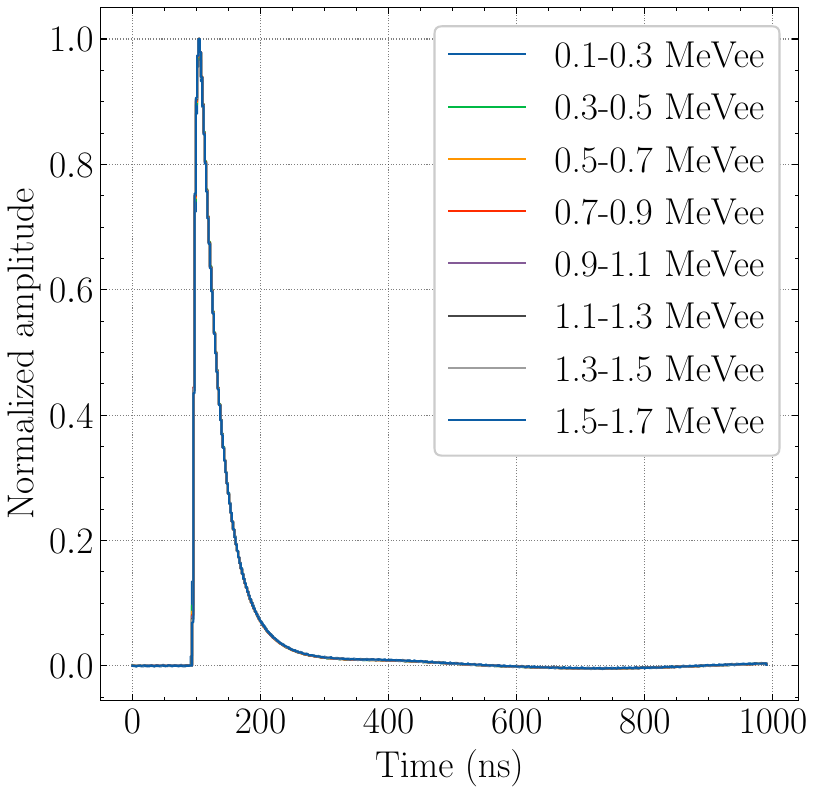}
        \caption{\added{Normalized average gamma-ray pulses.}}
        \label{fig:gamma_template_LO_binned_normalized}
    \end{subfigure}%
    \begin{subfigure}{.5\linewidth}
        \includegraphics[width=\textwidth]{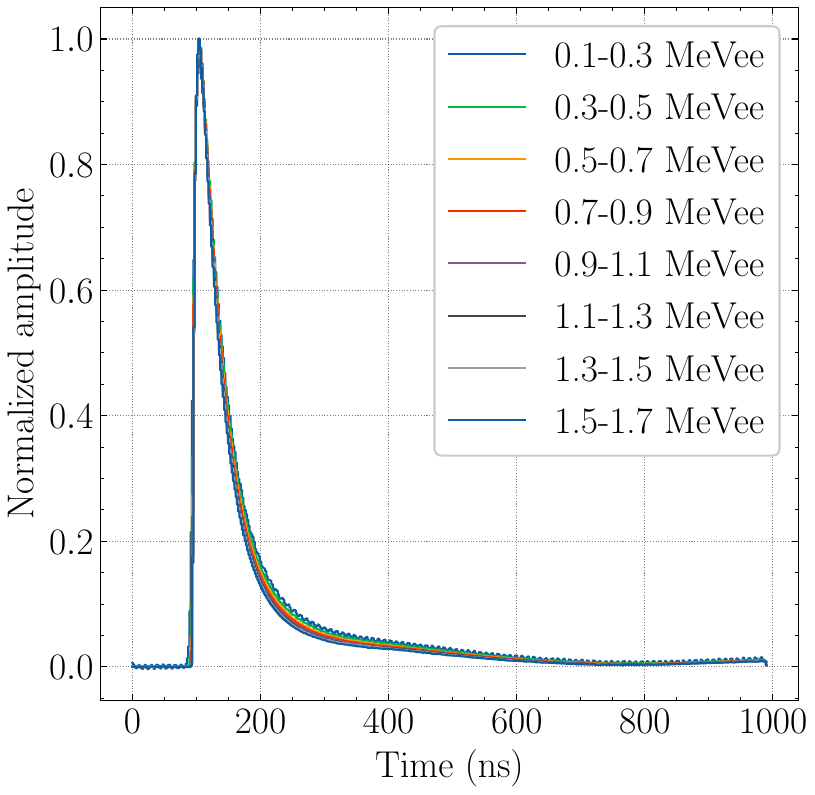} 
        \caption{\added{Normalized average neutron pulses.}}
        \label{fig:neutron_template_LO_binned_normalized}
    \end{subfigure}\\[1ex]
    \begin{subfigure}{\linewidth}
        \centering
        \includegraphics[width=.5\textwidth]{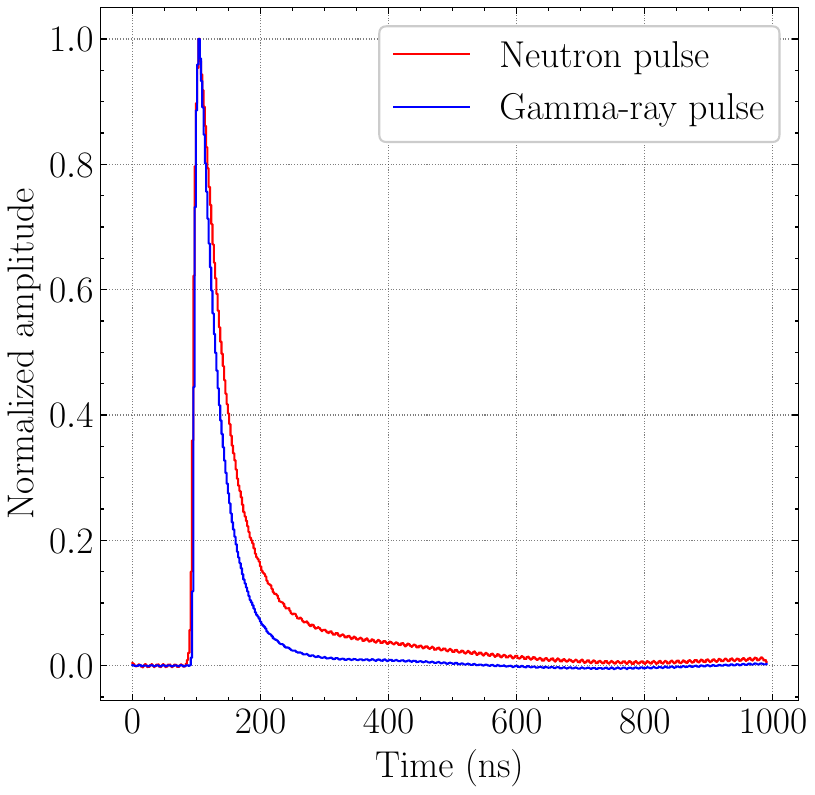}
        \caption{Templates created by averaging pulses in entire light output range.}
        \label{fig:template_whole_LO_range}
    \end{subfigure}
    \caption{{\replaced{Normalized neutron and gamma-ray template pulses acquired by measuring a $^{239}$PuBe source with a stilbene-d$_{12}$ crystal.}{Normalized deuterated stilbene neutron and gamma-ray \added{template }pulses.}}}
    \label{fig:dsb_pulses}
\end{figure}

PSD-capable scintillators emit blue-UV light upon interaction with ionizing radiation through prompt and delayed fluorescence~\cite{birks2013theory}. In 
organic scintillators, the light-emitting molecular excitation reactions are mainly due to recoil protons and electrons, produced by neutron and gamma-ray collisions, respectively. For an equivalent deposited energy, recoil protons produce a higher amount of delayed fluorescence,  compared to electrons, due to their higher ionization density, which results in a larger delayed component in the detected pulses~\cite{birks2013theory, Morishita2019207}. This difference in delayed components can then be used to discriminate between gamma-ray and neutron pulses. Examples of template neutron and gamma-ray pulses are shown in Figure~\ref{fig:template_whole_LO_range}, in which the neutron pulse shows a larger delayed component. \replaced{The template pulses were obtained by binning the measured pulses in 0.2 MeVee-wide bins from 0.1 MeVee to 1.7 MeVee shown by Figures~\ref{fig:gamma_template_LO_binned} and~\ref{fig:neutron_template_LO_binned}, normalizing the gamma-ray and neutron pulses to their maximum values shown by Figures~\ref{fig:gamma_template_LO_binned_normalized} and~\ref{fig:neutron_template_LO_binned_normalized} and then averaging them in each bin separately shown by Figure~\ref{fig:template_whole_LO_range}. The method used to obtain these pulses is described in further detail in Section~\ref{sec:dataset}.}{The method used to obtain these pulses is described in Section~\ref{sec:dataset} in greater detail.}

Traditionally, charge integration (CI)-based PSD \cite{POLACK2015253} is used to perform this discrimination. CI takes advantage of the difference in delayed components by defining the PSD parameter as the ratio of the area under the pulse tail to the total area (tail-to-total ratio) of the measured pulse, as shown by Equation~\eqref{eqn:CIparam}. The pulses are measured in voltages as designated by $V(t)$ and $t_0$ is the beginning of the pulse, $t_1$ is on the falling edge of the pulse, and $t_2$ is the end of the pulse.
\begin{equation}
    \mathrm{PSD_{CI}} = \frac{\sum_{t_1}^{t_2}V(t)}{\sum_{t_0}^{t_2}V(t)}
    \label{eqn:CIparam}
\end{equation}
Neutron pulses exhibit higher tail-to-total ratios due to the larger delayed components, which enables their discrimination from gamma-ray pulses.

\subsubsection{Digital Shaper PSD}
\begin{figure}[!htbp]
 \centering
  \includegraphics[width=1\linewidth]{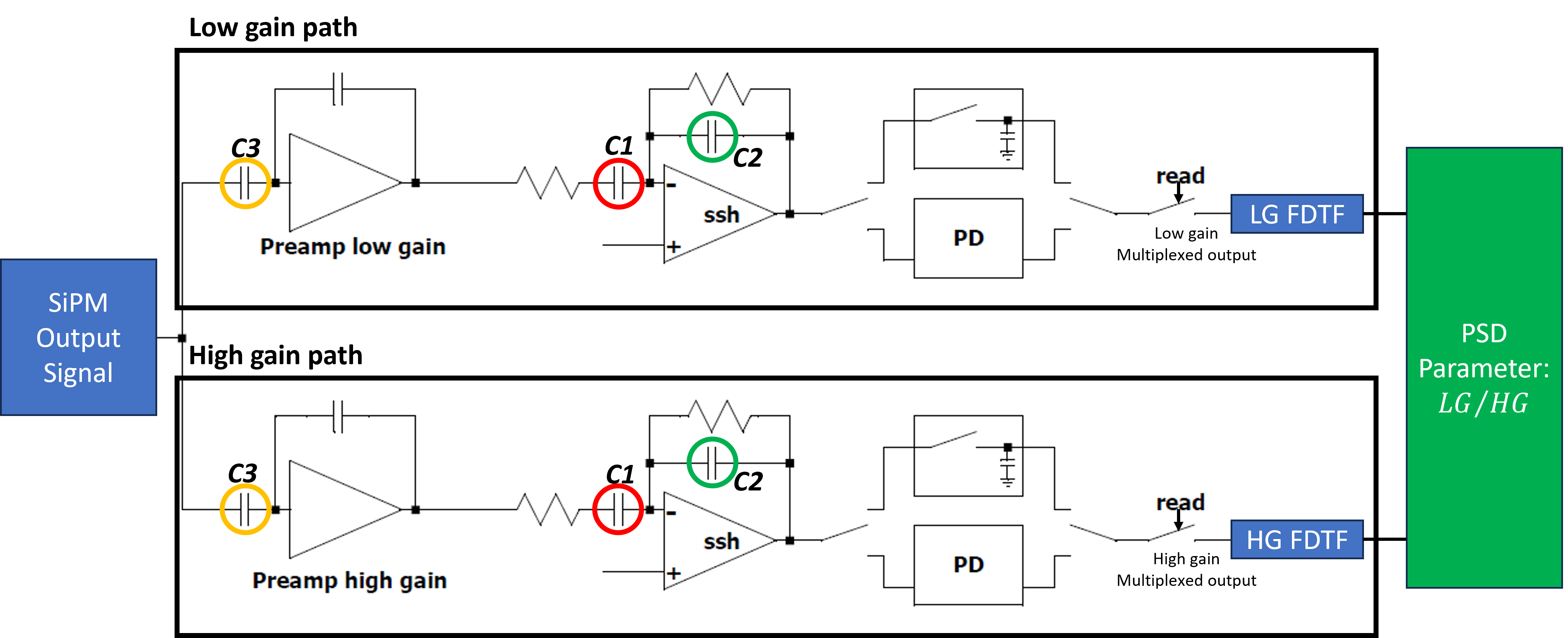}
  \caption{Calculation of the PSD parameter based on LG and HG outputs. SSH is the slow shaper and PD is the Peak Detector circuit~\cite{a55citx2020}.}
  \label{fig:psd_calculation}
\end{figure}

The Citiroc1A front-end encompasses a readout circuit with two independent paths, referred to as the low-gain (LG) and high-gain (HG) path. Each path consists of a low-gain/high-gain charge preamplifier, a pulse shaper, and a peak sensing circuit that detects and records the maximum of the shaped signal~\cite{a55citx2020}, as shown in Figure~\ref{fig:psd_calculation}. Since gamma rays and neutrons have different pulse shapes, they exhibit different responses to the shaper in the Citiroc1A.  It is therefore possible to perform PSD based on this difference in the shaper's responses.

 \begin{figure}[!htbp]
 \centering
  \includegraphics[width=\linewidth]{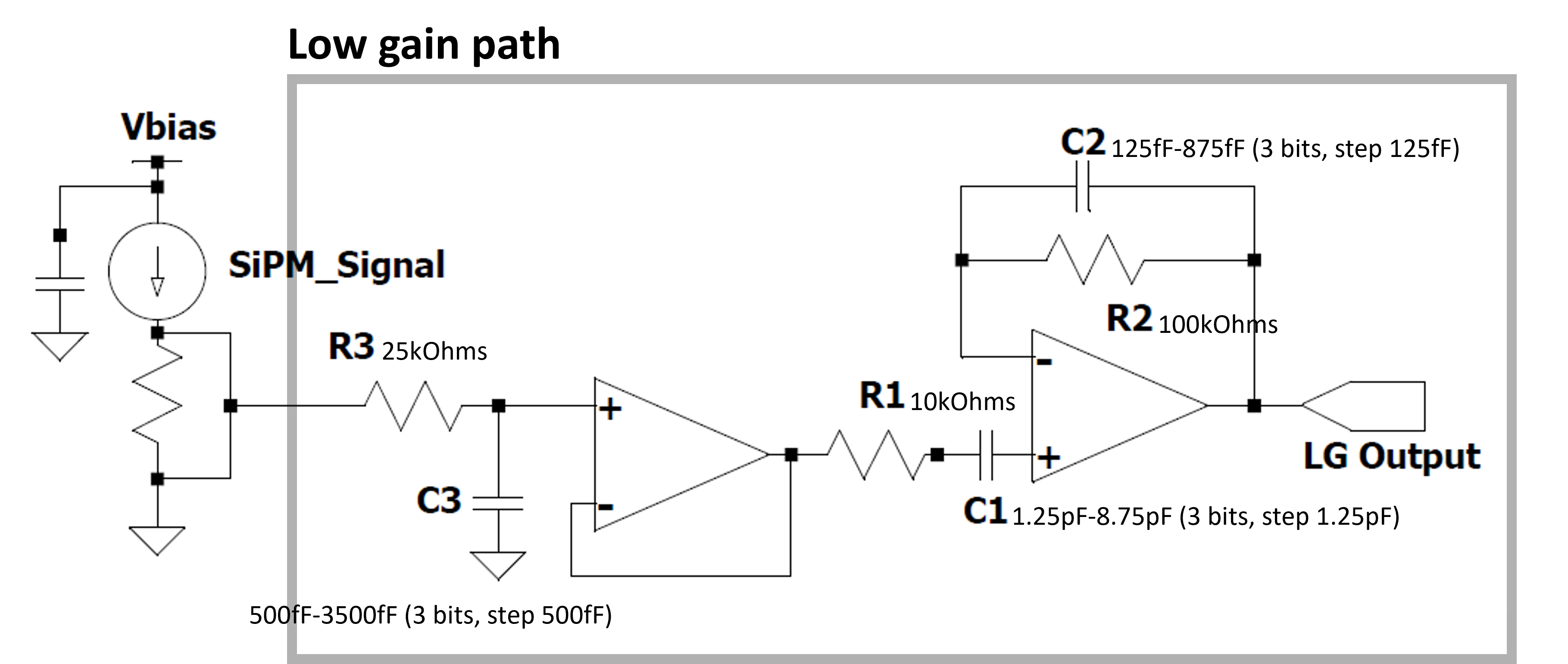}
  \caption{Single channel pulse shaper on the Citiroc1A. C1, C2, and C3 are the capacitances controlled by the peak time~\cite{citiroc1a}}
  \label{fig:pulse_shaper}
\end{figure}

For this method, the PSD parameter is defined as the ratio between the \replaced{HG and HG}{LG and HG} output signals as shown by Equation~\eqref{eqn:ASICparam}. 
Due to the two paths of the readout circuit being linear with energy, the PSD parameter is mainly a function of the pulse shape with a slight dependence on the pulse height~\cite{a5202}. Figure~\ref{fig:psd_calculation} illustrates the steps needed to calculate the PSD parameter. The pulse shaper is shown in Figure~\ref{fig:pulse_shaper}. Its response is dependent on a set of three adjustable capacitances - C\textsubscript{1}, C\textsubscript{2}, C\textsubscript{3} - also shown in Figure~\ref{fig:psd_calculation}. These capacitances are controlled through the ``peak time" setting in the readout software. There are a total of seven available peak times each associated with a specific set of capacitances.
To achieve the best PSD performance, we optimized the values of C\textsubscript{1}, C\textsubscript{2}, and C\textsubscript{3} to maximize the difference in PSD parameters of neutron and gamma-ray pulses. \added{A detailed description of the optimization method is in the following Section~\ref{sec:transferFunc}.}

\begin{equation}
    \mathrm{PSD_{ASIC}} = \frac{\mathrm{HG\ output}}{\mathrm{LG\ output}}
    \label{eqn:ASICparam}
\end{equation}

\subsection{\replaced{Computational Implementation of DS PSD}{Computational Methods}}\label{sec:sim_method}
In this section, we calculate the response of the ASIC's acquisition chain, based on which we can determine the optimal ASIC settings for PSD.


\subsubsection{Transfer Function and Pulse Shape}
\label{sec:transferFunc}
The response of the front-end acquisition chain of the ASIC was calculated based on the known architecture. A set of template gamma-ray and neutron pulses, shown in Figure~\ref{fig:dsb_pulses}, were used to test the response of the front-end to prototypical pulses.
The template neutron and gamma-ray pulses were obtained by averaging 10,000 neutron and gamma-ray pulses, respectively, and normalizing them to the single pulse maximum value. The method used to obtain these pulses is described in Section~\ref{sec:dataset} in greater detail.

We calculated the frequency-dependent-transfer-function (FDTF) of the pulse shaper (Figure~\ref{fig:pulse_shaper}) in the Laplace domain, shown by Equation~\eqref{eqn:FDTF}. $f$ stands for the frequency and $T(s)$ is the ratio between the output signal and input signal. C\textsubscript{1}, C\textsubscript{2}, and C\textsubscript{3} are the capacitances controlled by the peak time setting in peak sense mode of the readout software and the set relationship between the peak times and capacitances is shown in Table~\ref{table:peak_times}. 
The capacitances and R\textsubscript{1}, R\textsubscript{2}, and R\textsubscript{3} are shown in Figure~\ref{fig:pulse_shaper} as well. The resistive values are specific to the ASIC itself and cannot be tuned by the user via software.
\begin{equation}
\begin{gathered}
T(s)=\left(\frac{1}{s R_{3} C_{3}+1}\right)\left(\frac{s^{2} R_{1} C_{1} R_{2} C_{2}+s\left(R_{1} C_{1}+R_{2} C_{2}+R_{2} C_{1}\right)+1}{s^{2} R_{1} C_{1} R_{2} C_{2}+s\left(R_{1} C_{1}+R_{2} C_{2}\right)+1}\right)\\
s = i \omega, \omega = 2\pi f\label{eqn:FDTF}
\end{gathered}
\end{equation}

 \begin{table}[hbt!]
    \caption{Table of capacitances corresponding to the set peak time.}
    \label{table:peak_times}
    \centering
    \begin{tabular}{ |c|ccc| }
    \hline 
  Peak Time (ns) & C$_1$ (pF)  & C$_2$ (pF) & C$_3$ (pF) \\
    \hline 
    87.5 & 1.25& 125 & 500  \\
    75   & 2.50& 250 &1000    \\
    62.5 & 3.75&375  &1500  \\
    50   & 5.00&500  &2000 \\
    37.5 & 6.25& 625 &2500 \\
    25   & 7.50&750  &3000 \\
    12.5 & 8.75& 875 &3500  \\
    \hline 
    \end{tabular}
\end{table}

\begin{figure}[!htbp]
 \centering
  \includegraphics[width=.8\linewidth]{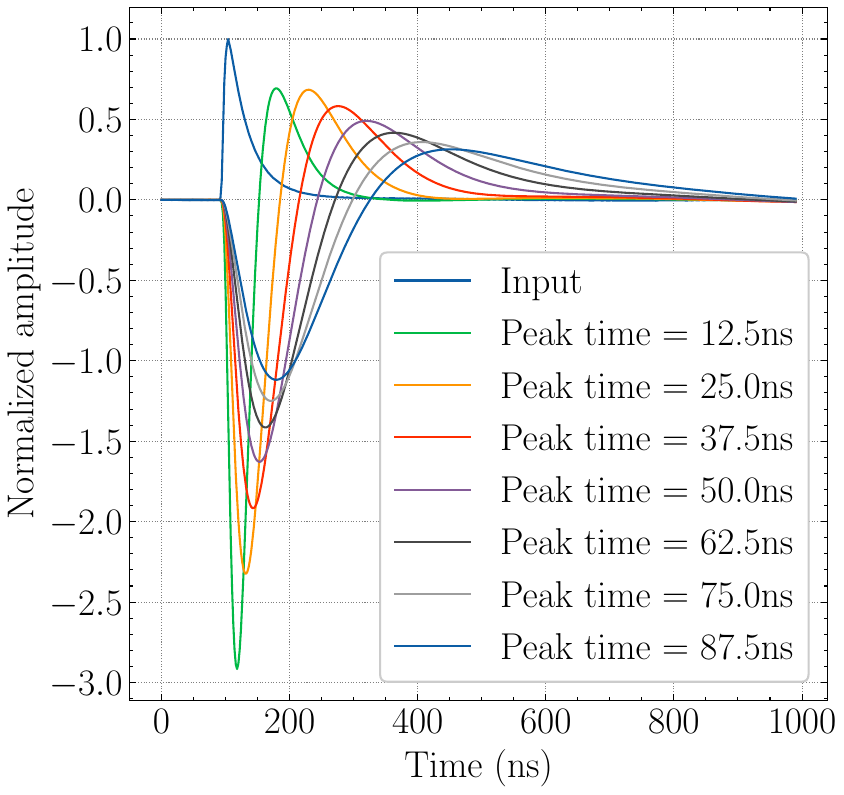}
  \caption{Signals output by FDTF with different peak times.}
  \label{fig:filter_pulses}
\end{figure}

We used the \added{normalized} template gamma-ray pulses in Figure~\ref{fig:template_whole_LO_range} as the input to the FDTF to determine how the ``peak time'' affects the shape of the output signal. Figure~\ref{fig:filter_pulses} shows the\added{ unnormalized} shape of the output signal at different peak times. 
As expected, a longer peak time corresponds to an increased delay of the peak timestamp of the shaped pulse.

\subsubsection{FDTF Optimization}

The user can set the seven peak time values independently for the LG and HG paths, corresponding to 49 different combinations. For every combination of LG and HG peak times, we feed a template pulse to the LG and HG transfer functions, calculate the LG and HG peak heights, and finally calculate the PSD parameter, which is the ratio between the \replaced{HG and LG}{LG and HG} output signals.
This procedure is performed for both gamma-ray and neutron template pulses. The combination of peak times that yields the largest difference between the gamma ray and neutron PSD parameters is optimal for discriminating the two types of pulses and was hence selected in this study. 
 


\subsection{Experimental Methods}\label{sec:dataset}
With the optimized PSD settings, i.e., peak times, the DS PSD method is tested using two different data sets, one from a pulse generator and the other from measured data. The pulse generator was used to explore a broader range of pulse shapes and the measured data was used to determine the performance of the DS PSD.
\subsubsection{Data Sets Used to Test the PSD performance}\label{sec:data_acq}
\begin{figure}[!htbp]
 \centering
  \includegraphics[width=\linewidth]{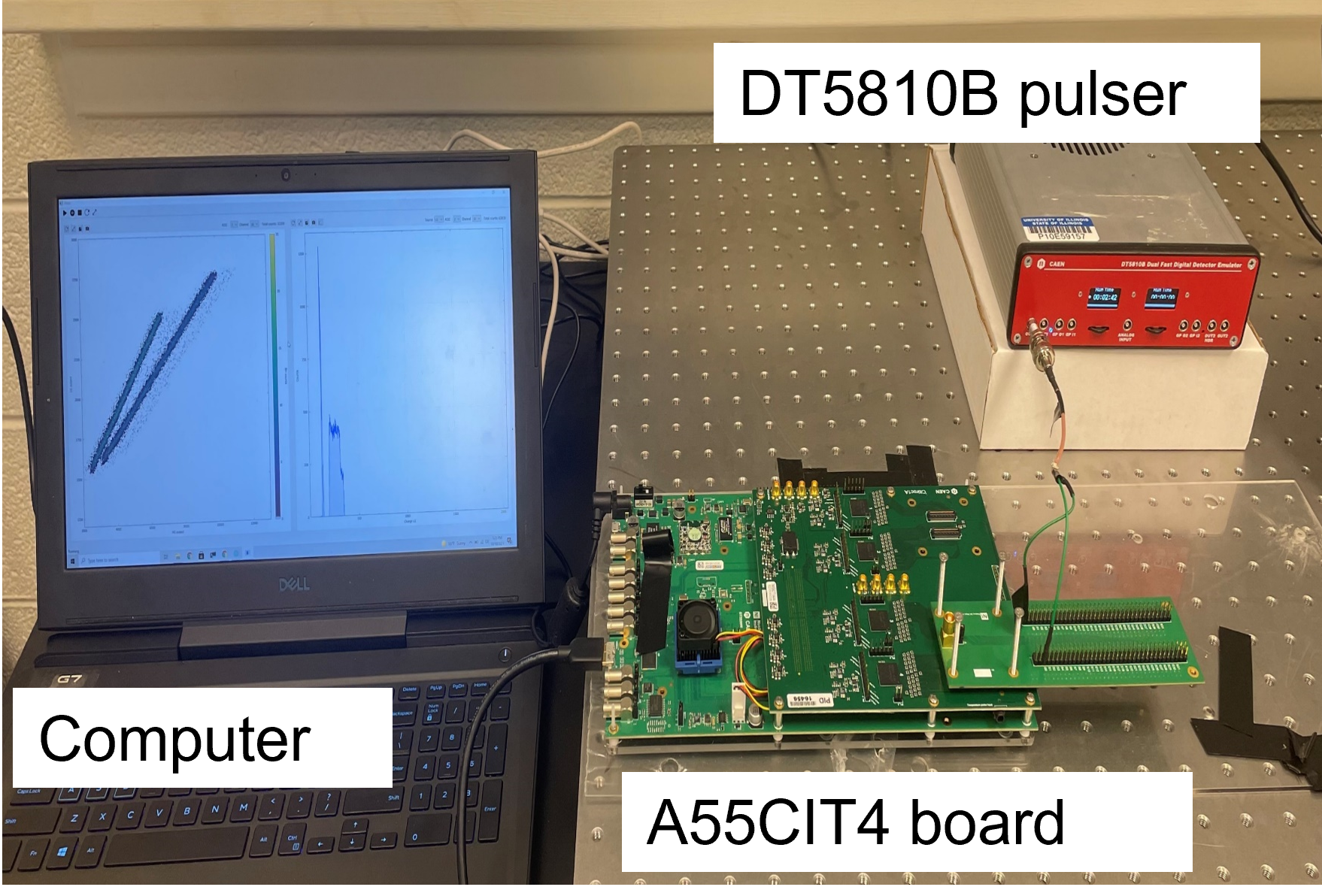}
  \caption{Emulated pulses experimental setup. The pulse generator output is connected directly to the ASIC and the output is recorded and processed in real-time by the computer to perform PSD.}
  \label{fig:emulator_exp_setup}
\end{figure}

\begin{figure}[!htbp]
 \centering
  \includegraphics[width=.8\linewidth]{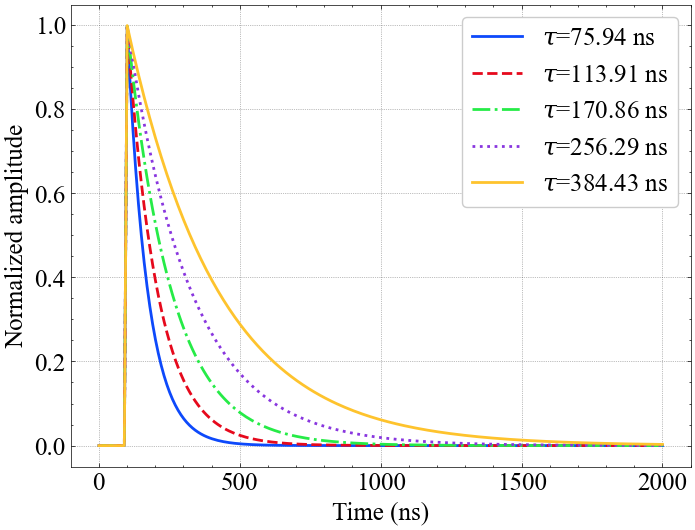}
  \caption{Normalized measured emulator pulses.}
  \label{fig:emulator_pulses}
\end{figure}

The emulated pulses were generated using the experimental setup in Figure~\ref{fig:emulator_exp_setup}. A DT5810B pulse generator~\cite{dt5810B} was directly connected to the A55CIT4 board. \replaced{A laptop was connected to the A55CIT4 ASIC to program the board and acquired and visualize the PSD processed data.}{ The board was connected to a laptop where PSD was performed in real time.} The pulse generator created exponential pulses with a pulse shape defined by Equation~\eqref{eqn:PulseGenExponentialEq}, where $\tau$ is the decay constant and OFFSET is \SI{100}{\ns}. 
\begin{equation}
A\times
    \begin{cases}
        e^{\frac{-(t-\mathrm{OFFSET})}{\tau}} & t \geq \mathrm{OFFSET} \\ 
        0 & \mathrm{otherwise}
    \end{cases}
    \label{eqn:PulseGenExponentialEq}
\end{equation}
Five decay constants were tested and the pulse height was uniformly distributed between 0 and approximately 0.3V. The shapes of the emulated pulses are shown in Figure~\ref{fig:emulator_pulses}.


\begin{figure}[!htbp]
 \centering
  \includegraphics[width=\linewidth]{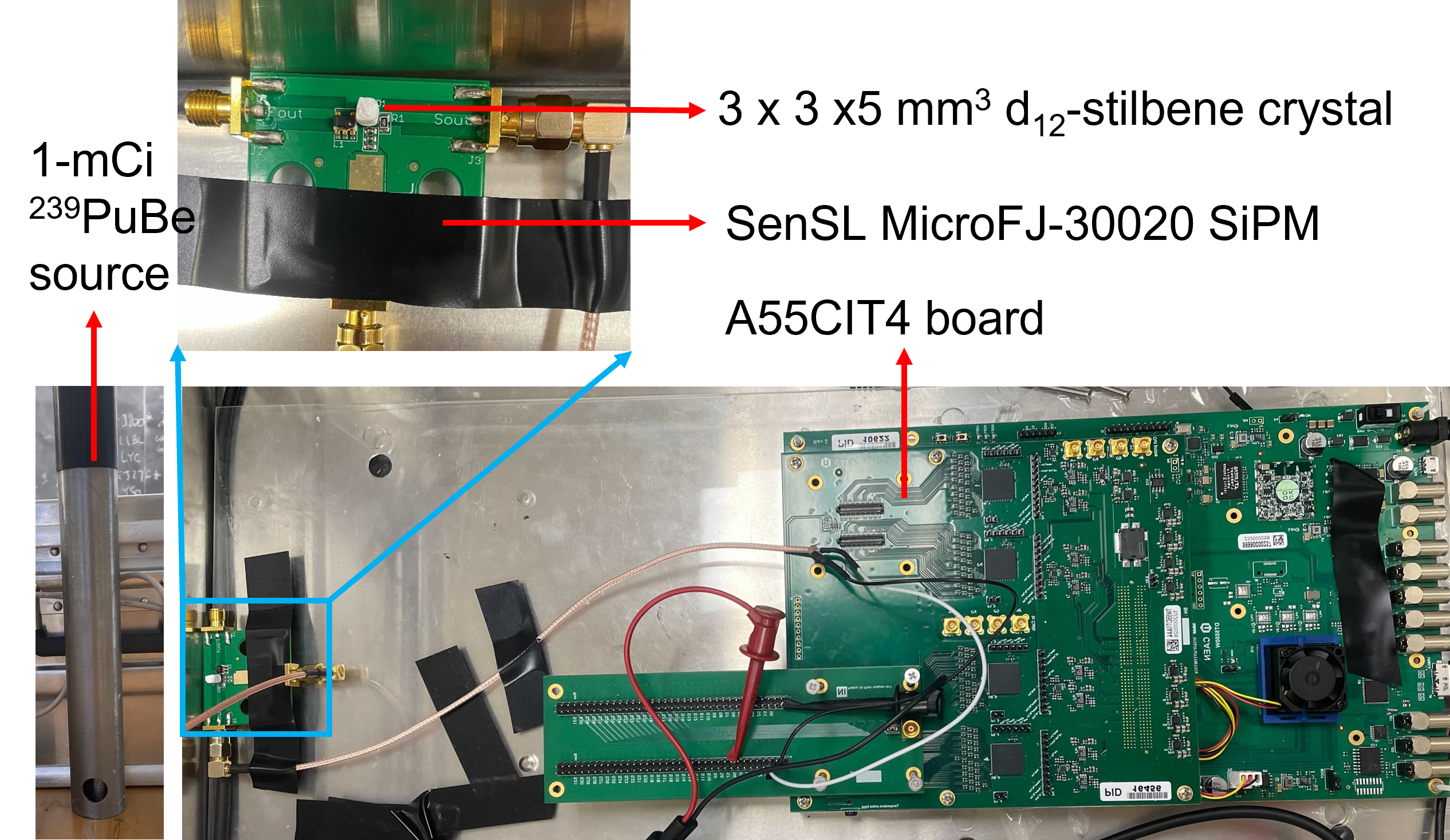}
  \caption{Measured data experimental setup. The SiPM output is sent to the ASIC and the signal is recorded by a computer.}
  \label{fig:dsb_exp_setup}
\end{figure}

The experimentally measured data set was obtained using a $3\times3\times5$\SI{}{\cubic\mm} deuterated stilbene (stilbene-d\textsubscript{12}) crystal coupled to a SenSL MicroFJ-30020 SiPM~\cite{sensljseries2015}. \added{We set the over-voltage of MicroFJ-30020 SiPM to \SI{2}{V} and the max signal amplitude was \SI{0.2}{V}. The signal amplitude was low enough so that no severe signal saturation occurred in the HG channel, and high enough so that we could still acquire low-amplitude signals in the low light output range.} The setup of DS PSD is shown in Figure~\ref{fig:dsb_exp_setup}. The SiPM signal is processed by the A55CIT4 board. A 1-mCi \replaced{$^{239}$}{\textsubscript{239}}PuBe source was placed at a 10-cm distance from the deuterated stilbene crystal and measured for 30 minutes, resulting in approximately 200,000 pulses. For comparison, we replaced the A55CIT4 board with a 14-bit 500-MSps DT5730 digitizer and performed CI PSD on the saved pulses. The data set acquired with the digitizer was also used for the optimization of peak times described in Section~\ref{sec:sim_method}. Examples of measured pulses\deleted{, normalized to the single pulse maximum,} are shown in Figure~\ref{fig:dsb_pulses}. 

\subsubsection{Data Processing and Figure-of-Merit}
\label{sec:DataProcessing}
PSD is performed on the emulated data and experimentally measured data using both the traditional charge integration method and the new PSD algorithm explained in the Section~\ref{sec:PSD_Method}. The CI parameters for the total integral and tail-to-total integral are \SI{300}{\ns} starting \SI{50}{\ns} before the trigger and \SI{220}{\ns} starting \SI{30}{\ns} after the trigger, respectively.
The results of the DS PSD are then compared with the results of the charge integration PSD. This comparison includes the PSD plots for a qualitative comparison and the calculated figure of merit (FOM) for a quantitative comparison.
The FOM is calculated by performing a double Gaussian fit of the distribution of PSD parameters using Equation~\eqref{eqn:double_gauss} and then using the parameters of the Gaussian fit as inputs to the FOM equation (Equation~\eqref{eqn:FOM})~\cite{7876820}. $\mu$ is the mean of one peak and FWHM\added{=$2\sqrt{2\ln2}\sigma$} is the full-width half-maximum of the same peak. A higher FOM is preferable because it corresponds to a wider separation between the neutron and gamma ray pulses in the PSD-parameter space.
\begin{equation}
    f(x) = a_n \exp\Big(-\frac{(x-\mu_n)^2}{2\sigma_n^2}\Big) + a_{\gamma} \exp\Big(-\frac{(x-\mu_{\gamma})^2}{2\sigma_{\gamma}^2}\Big)
    \label{eqn:double_gauss}
\end{equation}
\begin{equation}
\begin{gathered}
\mathrm{FOM} = \frac{|\mu_n - \mu_{\gamma}|}{\mathrm{FWHM}_n + \mathrm{FWHM}_{\gamma}}
\end{gathered}
\label{eqn:FOM}
\end{equation}

\section{Results}\label{sec:results}
In this section we present the results of the optimization of the peak times based on the simulation of the circuit and the application of the DS PSD to emulated pulses as well as deuterated stilbene pulses.
\subsection{FDTF Optimization}
\begin{figure}[!htbp]
 \centering
  \includegraphics[width=.8\linewidth]{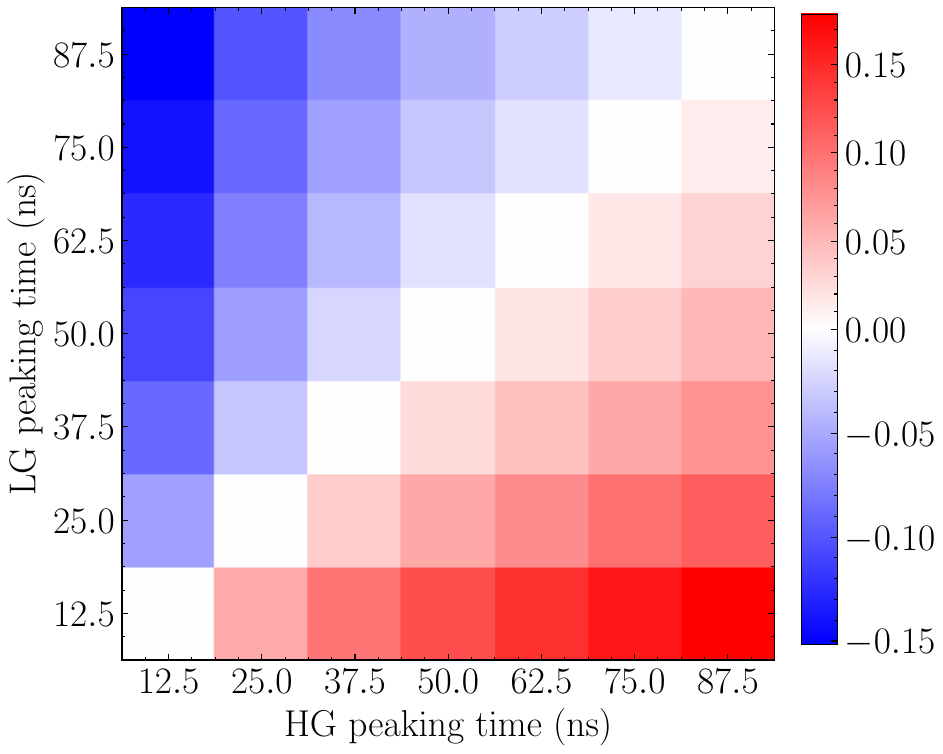}
  \caption{$\frac{PSD_n - PSD_{\gamma}}{PSD_{\gamma}}$ as a function of LG peaking time and HG peak time. \added{The combination of 87.5 ns for the LG peaking time and 12.5 ns for the HG peaking time, produces the largest difference between the LG and HG PSD parameters than any other combination. This is shown by the colorbar labels as the darkest red is greater than 0.15 whereas the darkest blue is exactly -0.15.}}
  \label{fig:optimizing_peak_time}
\end{figure}

\added{We optimized the FDTF using the template pulses in Figure~\ref{fig:template_whole_LO_range} as the input.}
Figure~\ref{fig:optimizing_peak_time} shows the relative difference between neutron and gamma-ray PSD parameters for all 49 possible combinations of LG and HG peak times.
The x-axis is the LG peaking time and the y-axis is the HG peaking time. Each small square is colored by the relative difference between the neutron and gamma-ray PSD parameter. 
The optimization showed that the difference between the LG and HG peaking times should be maximized in order to maximize the difference in PSD parameters. The combination that maximizes the difference between the LG and HG output values is 12.5 ns and 87.5 ns for the LG and HG peaking times, respectively. These optimized settings were used in the following experiments.

\subsection{Computational  Results}
\begin{figure}[!htbp]
    \centering
    \includegraphics[width=\linewidth]{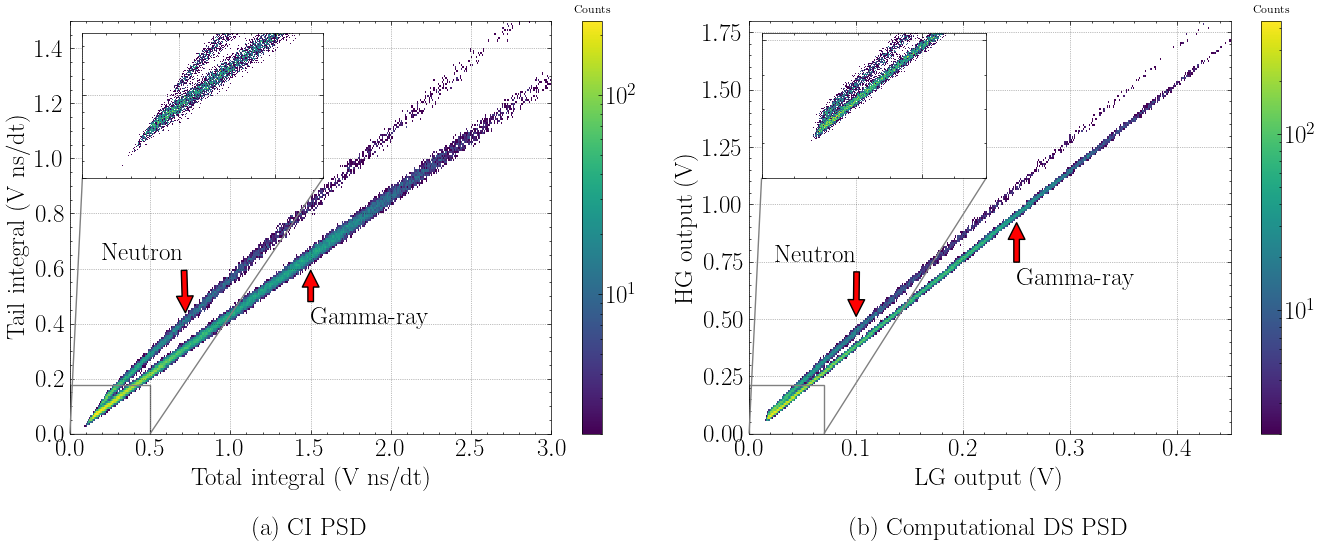}
    \caption{Comparison of (a) charge integration PSD and (b) computational DS PSD of the measured dataset\added{ acquired by measuring a PuBe with a stilbene-d$_{12}$ crystal}. }
    \label{fig:compare_psd_dataset}
\end{figure}

With the optimized settings known, we can now test the performance of the DS PSD on \replaced{dataset C}{our measured dataset}\added{, described in Section ~\ref{sec:data_acq}}. The results of charge integration PSD\added{, acquired using dataset B,} and \added{computational} DS PSD results are shown in Figure~\ref{fig:compare_psd_dataset} for comparison. Figure~\ref{fig:compare_psd_dataset}a shows the charge integration result where the x-axis is the total pulse integral and the y-axis is the tail integral. 
Figure~\ref{fig:compare_psd_dataset}b shows the PSD obtained with the \added{computational} DS PSD method. The x-axis is the peak height of the \replaced{LG}{HG} signal and the y-axis is the peak height of the \replaced{HG}{LG} signal. We observed a good separation between the two groups of pulses, showing the feasibility of DS PSD on this dataset. \replaced{Figure~\ref{fig:compare_FOM_dataset} shows the calculated FOMs for all the pulses shown in Figure~\ref{fig:compare_psd_dataset}, which range from 50 keVee to 1200 keVee.}{Figure~\ref{fig:compare_FOM_dataset} shows the calculated FOMs from Figure~\ref{fig:compare_psd_dataset}. }



\begin{figure}[!htbp]
    \centering
    \includegraphics[width=\linewidth]{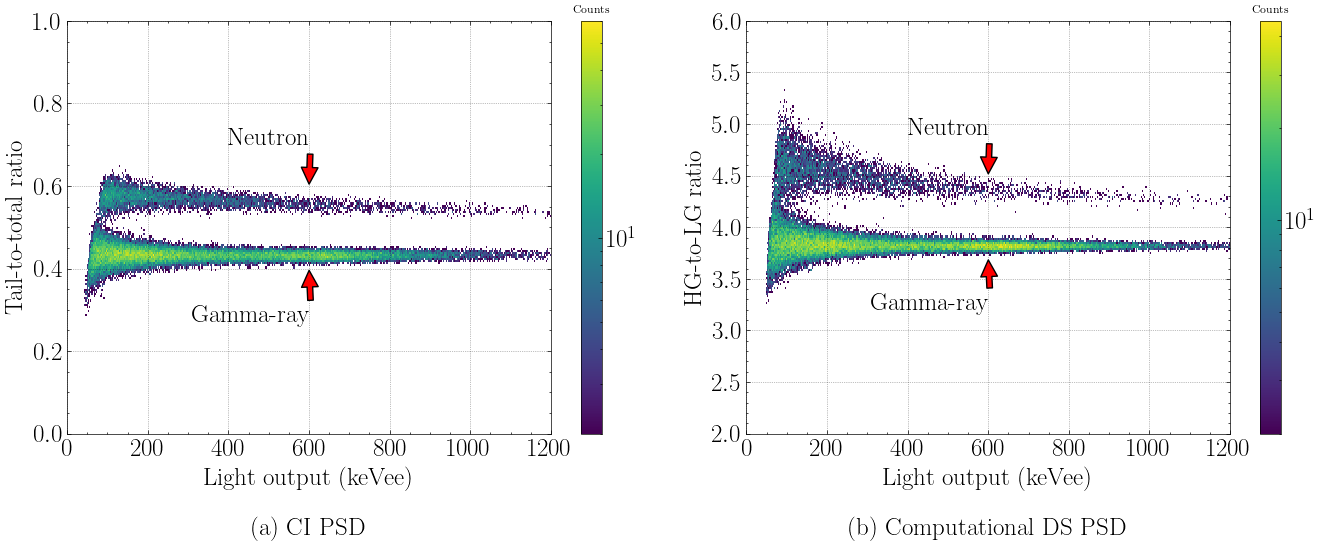}
    \caption{\added{Comparison of (a) charge integration PSD and (b) computational DS PSD of the measured dataset acquired by measuring a PuBe with a stilbene-d$_{12}$ crystal}. }
    \label{fig:compare_ratio_dataset}
\end{figure}

\begin{figure}[!htbp]
    \centering
    \includegraphics[width=\linewidth]{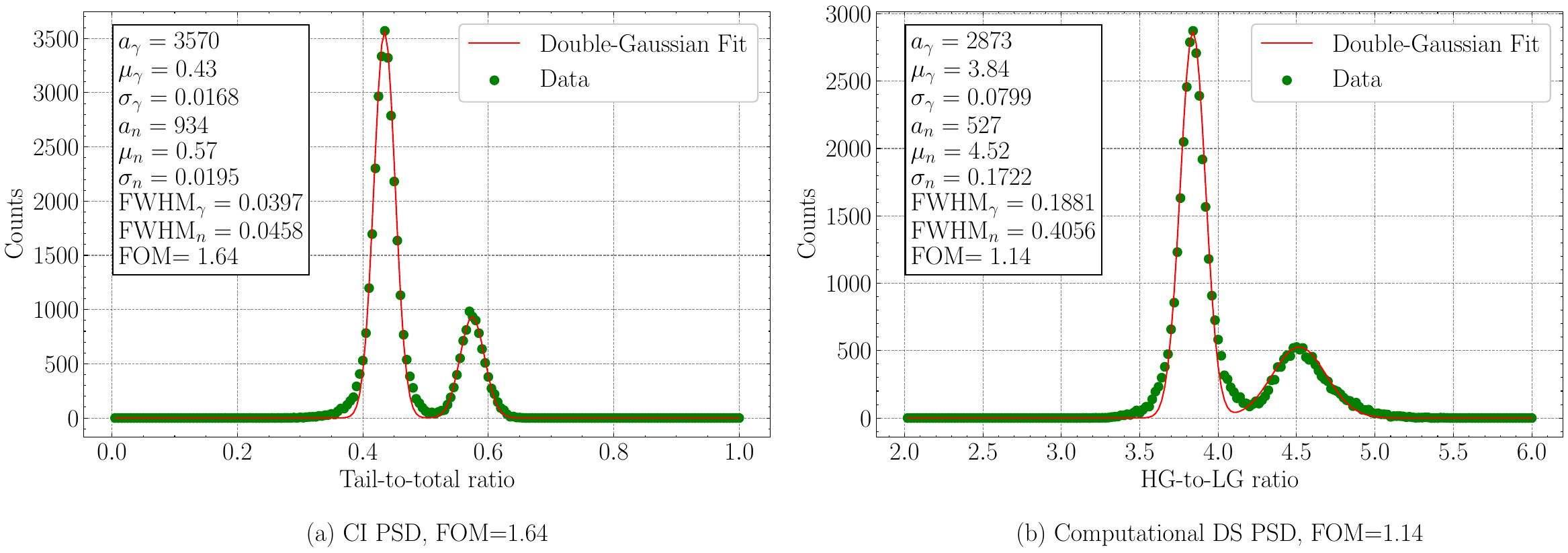}
    \caption{Comparison of the FOM for (a) charge integration PSD and (b) computational DS PSD of the measured dataset\added{ acquired by measuring a PuBe with a stilbene-d$_{12}$ crystal}.}
    \label{fig:compare_FOM_dataset}
\end{figure}

\subsection{Experimental Results}
\begin{figure}[!htbp]
 \centering
  \includegraphics[width=\linewidth]{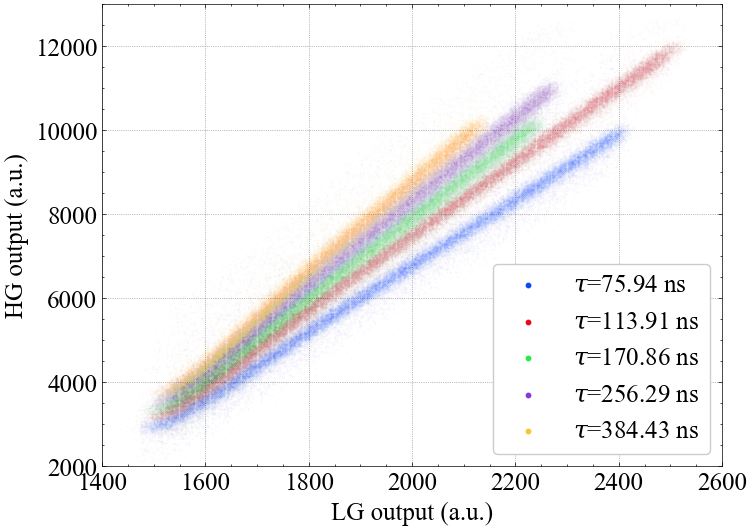}
  \caption{DS PSD of emulated pulses with different time constants.}
  \label{fig:asci_psd_emulator}
\end{figure}

We \replaced{used dataset A}{generated emulated pulses with five different decay constants}, described in Sec.~\ref{sec:data_acq}, and input them to the A55CIT4-DT5550W system. We recorded the LG and HG output for each pulse and created a scatter plot of the LG output as a function of the HG output shown in Figure~\ref{fig:asci_psd_emulator}. Each color represents one pulse type with a different time constant. The x-axis and y-axis are the \replaced{LG}{HG} output and \replaced{HG}{LG} output in ADC units, respectively. The distinct separation shows that emulated pulses with different time constants yield a different PSD parameter and can be easily discriminated.

\begin{figure}[!htbp]
    \centering
    \includegraphics[width=\linewidth]{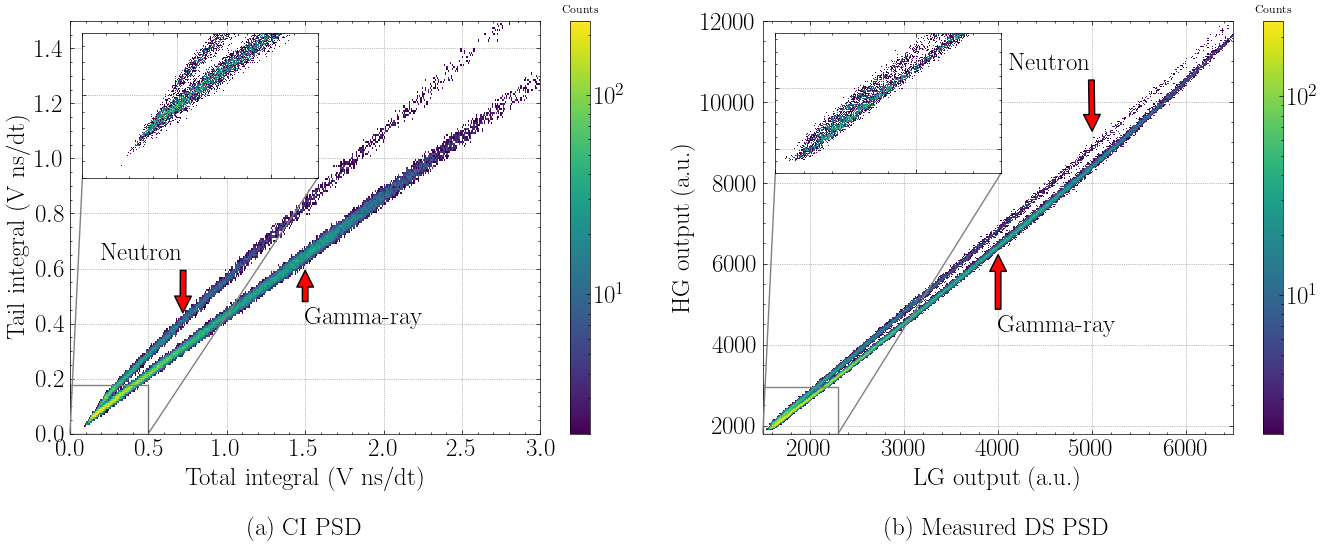}
    \caption{Comparison of (a) charge integration PSD and (b) DS PSD of deuterated stilbene pulses\added{ acquired by measuring a PuBe with a stilbene-d$_{12}$ crystal}.}
    \label{fig:compare_psd_dsb}
\end{figure}

The same analysis was performed on \replaced{datasets B and C}{the measured deuterated stilbene dataset}\added{, described in Section ~\ref{sec:data_acq}}. The results of charge integration PSD are shown in Figure~\ref{fig:compare_psd_dsb}a. The x-axis and y-axis are the total and tail integral, respectively. Figure~\ref{fig:compare_psd_dsb}b shows the \added{physical implementation of }DS PSD. The x-axis and y-axis are the \replaced{LG}{HG} output and \replaced{HG}{LG} output, respectively. Separation between the neutron and gamma-ray bands is observed, indicating that DS PSD allows us to discriminate between neutrons and gamma rays. 
Using the method described in Section~\ref{sec:DataProcessing}, we calculated the FOM for both the DS PSD and CI PSD methods, as shown by Figure~\ref{fig:FOM}. \added{We also calculate the FOM for the first three light output bins for the CI and DS PSD methods and show the relationship between the PSD performance and light output in Figure~\ref{fig:FOM_LO_binned}. One can observe in Figure 17b (150-200 keVee) allows us to discriminate between gamma-ray and neutron pulses with minimum overlap between the two Gaussian distributions corresponding to these radiation types.} The FOM of the DS PSD is smaller than that of the charge-integration.  


\begin{figure}[!htbp]
    \centering
    \includegraphics[width=\linewidth]{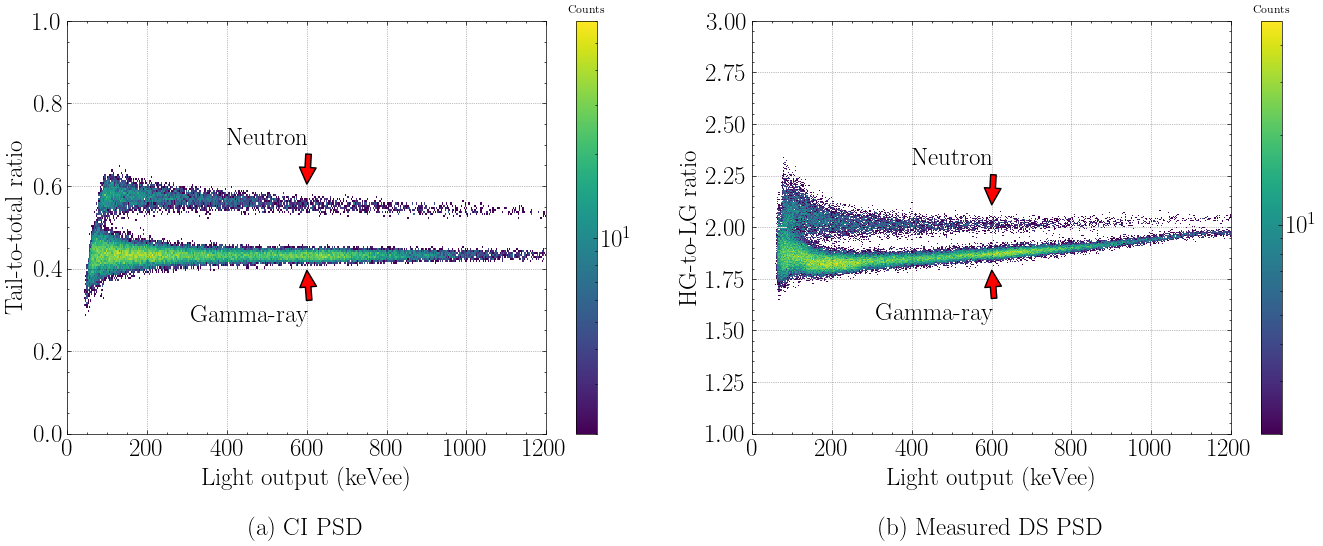}
    \caption{\added{Comparison of (a) charge integration PSD and (b) DS PSD of deuterated stilbene pulses acquired by measuring a PuBe with a stilbene-d$_{12}$ crystal}.}
    \label{fig:compare_ratio_dsb}
\end{figure}




\begin{figure}[!htbp]
    \centering
    \includegraphics[width=\linewidth]{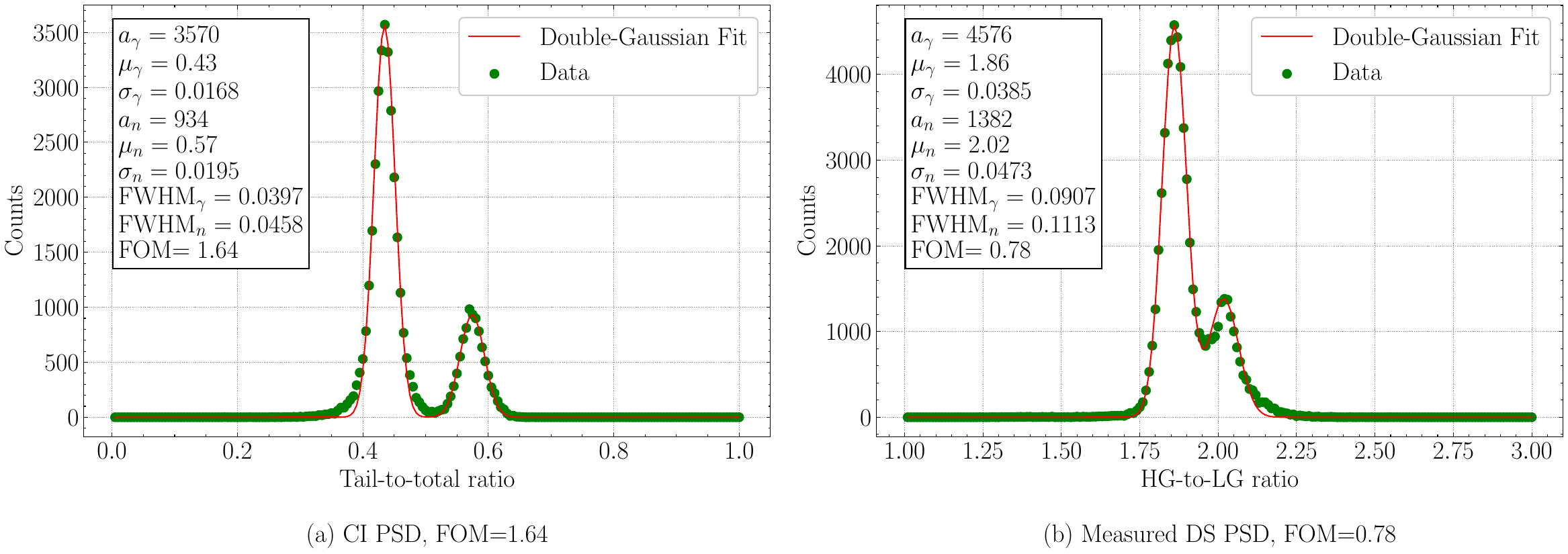}
    \caption{Comparison of the FOM for (a) charge integration PSD and (b) DS PSD\added{ acquired by measuring a \textsuperscript{239}PuBe with a stilbene-d$_{12}$ crystal}.}
    \label{fig:FOM}
\end{figure}

\begin{figure}
    \centering
    \begin{subfigure}{\textwidth}
        \includegraphics[width=\textwidth]{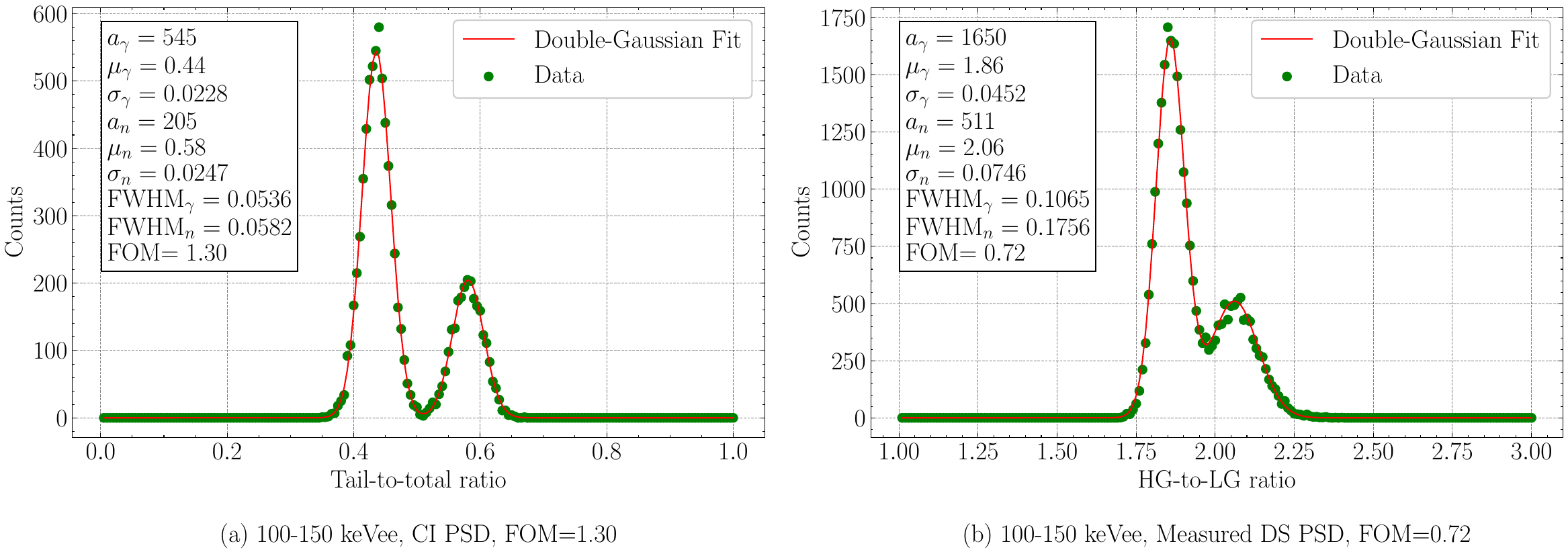}
        \label{fig:FOM_100_150_keVee}
    \end{subfigure}\\[1ex]
    \begin{subfigure}{\linewidth}
        \includegraphics[width=\textwidth]{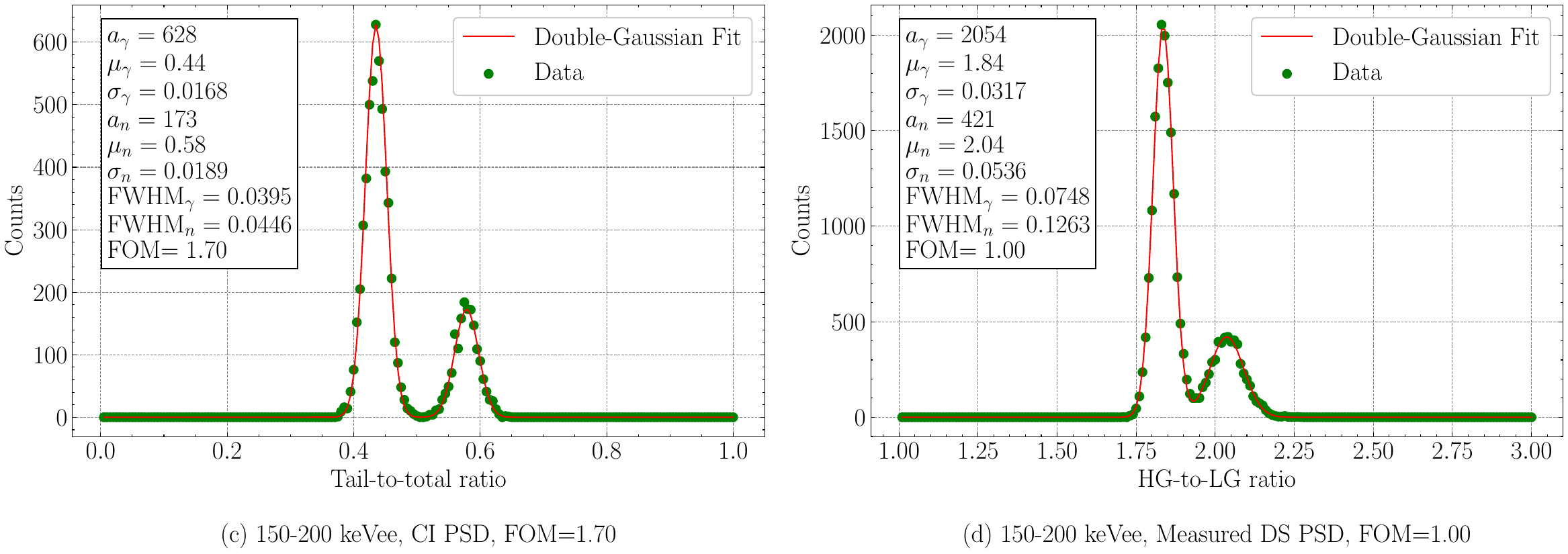}
        \label{fig:FOM_150_200_keVee}
    \end{subfigure}\\[1ex]
    \begin{subfigure}{\linewidth}
        \centering
        \includegraphics[width=\textwidth]{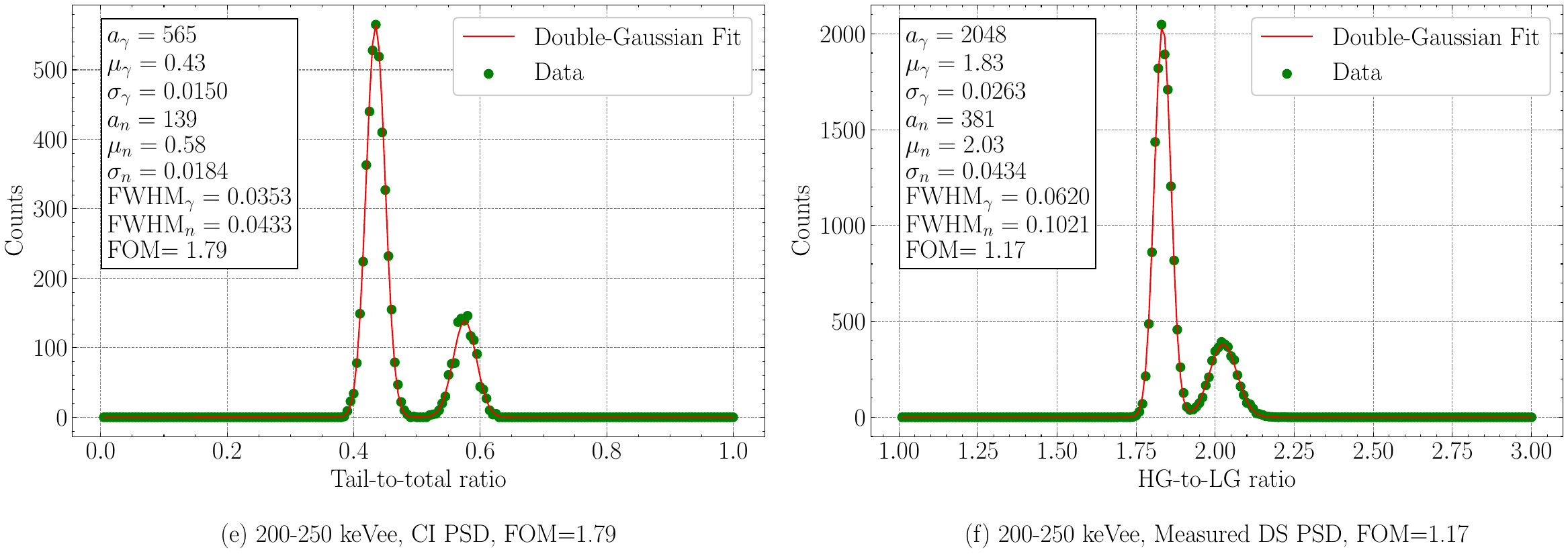}
        \label{fig:FOM_200_250_keVee}
    \end{subfigure}
    \caption{\added{PSD performance in the first three light output bins.}}
    \label{fig:FOM_LO_binned}
\end{figure}


\section{Discussion and Conclusions}\label{sec:conclusion}
An on-board ASIC-based PSD method was developed on a commercially available platform, i.e., the Citiroc1A. This DS PSD method enables the discrimination of gamma-ray and neutron pulses within a light output range of 0.1 MeVee to 2 MeVee. The presented results show that the commercially available ASIC-based readout is capable of performing on-board PSD. Unlike the traditional CI method, we used the LG and HG paths intrinsic to the Citiroc1A to extract a shape-dependent PSD parameter, defined as the ratio between the output signal of the LG and HG paths. We developed an FDFT model to emulate the response of the shaping circuit in each path and then, using the FDFT model, we optimized the PSD parameter, i.e., the peak times of the LG and HG front end. The peak times that provided the best PSD parameter are 12.5-ns for the LG path and 87.5-ns for the HG path. The optimized PSD performance was estimated and compared to CI using a template-pulse data set. This study showed that the DS PSD performance is comparable to the current CI PSD. We then applied the DS PSD to synthetic pulses generated with a pulse emulator with controlled decay constants and confirmed experimentally that DS PSD can discriminate pulses of varying shapes. Finally, we applied the DS PSD to pulses acquired by measuring a $^{239}$PuBe source with a deuterated stilbene crystal coupled to a SiPM. 
For this data set, the calculated FOM for DS PSD is 0.78 and 1.64 for CI PSD. \replaced{Overall CI PSD outperforms DS PSD by 50\%. Out of this 50\%, about 30\% can be attributed to the limited number of shaping times. This is shown by performing the computational analysis of DS PSD using teh measured data set. Th remaining observed decrease in FOM for the physical implementation of the DS PSD is likely due to noise on the low gain and high gain-shaped output signals, intrinsic to the ASIC itself. Although the DS PSD performance is worse than that of CI PSD, neutron and gamma-ray pulses can be effectively discriminated at light output values higher than \SI{0.15}{\MeVee}, corresponding to a neutron deposited energy of approximately \SI{0.75}{\MeV}.  }{Although the FOM for DS PSD is 50\% lower than that of the CI PSD, neutron and gamma-ray pulses can be effectively discriminated at light output values higher than \SI{0.15}{\MeVee}, corresponding to a neutron deposited energy of approximately \SI{0.75}{\MeV}.} It should also be noted that DS PSD is performed on-board, in real time, and does not require the transfer of data to a separate system for further processing. \deleted{We noticed a degradation of PSD performance of the DS PSD method when comparing the on board implementation and the calculation of the FDTF in python. This loss in performance could be attributed to the effect of noise on the low gain and high gain-shaped output signals.}
\replaced{T}{It should be noted that t}he Citiroc1A in this work has a total of seven available peak times. The DS PSD method could potentially be improved by finer tuning of the peak time, rather than selecting one of the seven options provided. \added{We plan on testing the DS PSD performance using different organic scintillators, such as EJ-276D and small-molecule organic glass.}

\section*{Acknowledgments}
This material is based upon work supported by the Department of
Energy National Nuclear Security Administration through the Nuclear Science and Security Consortium under Award Number DE-NA0003996.
\bibliography{references}

\end{document}